\documentstyle[proceedings,numreferences,epsfig]{crckapb} 
\newcommand{\stt}{\small\tt}
\newcommand{\DD}{\mbox{$\cal D$}}
\newcommand{\GG}{\mbox{$\cal G$}}
\newcommand{\SS}{\mbox{$\cal S$}}
\def\laq{\raise 0.4ex\hbox{$<$}\kern -0.8em\lower 0.62
ex\hbox{$\sim$}}
\def\be{\begin{equation}}
\def\ee{\end{equation}}
\def\bea{\begin{eqnarray}}
\def\eea{\end{eqnarray}}

\def\bk {{\bf k}}
\begin{opening}
\title{THE ROLE OF TOPOLOGICAL DEFECTS IN COSMOLOGY}

\author{M. SAKELLARIADOU}
\institute{Division of Astrophysics, Astronomy, and Mechanics\\
           Department of Physics, University of Athens\\
	   Panepistimiopolis, GR-15784 Zografos (Athina), Hellas}
\end{opening}

\begin{document}

{\stt abstract} Topological defects are involved in a plethora of
physical and astrophysical phenomena.  In these lectures, I will
review the r\^ ole they could play in the large-scale structure
formation and the anisotropies of the cosmic microwave background, as
well as in various high energy phenomena, including baryon number
asymmetry, ultra-high energy cosmic rays, and gamma ray bursts. I will
then summarize the gravitational effects of cosmic strings. Finally, I
will briefly discuss the r\^ ole of topological defects in brane world
cosmology.
   
\section{Introduction}
Most aspects of high energy physics beyond the standard model can only
be tested by going to very high energies, which are by far greater
than those accessible by present, or even future, terrestrial
accelerators.  Cosmology has offered a way to {\sl experimentally} test new
theories of fundamental forces. In these lectures, I will shortly
present the fascinating interplay between particle physics and
cosmology, as provided by topological defects.

\par Many particle physics models of matter admit solutions which
correspond to a class of topological defects, that are either stable
or long-lived.  Provided our understanding about unification of forces
and the big bang cosmology are correct, it is natural to expect that
such topological defects could have formed naturally during phase
transitions followed by spontaneously broken symmetries, in the early
stages of the evolution of the universe.  Certain types of defects
lead to disastrous consequences for cosmology and thus, they are
undesired, while others may play a useful r\^ole.

\par Spontaneous symmetry breaking is an old idea, having its origin
in the field of condensed matter physics, where it is described in
terms of the order parameter, while in particle physics, symmetry
breaking is described in terms of a scalar field, the Higgs field.
The symmetry is called spontaneously broken (SSB) if the ground state
is not invariant under the full symmetry of the Lagrangian
density. Thus, the vacuum expectation value of the Higgs field is
non-zero. In quantum field theories, broken symmetries are restored at
high enough temperatures, as one can easily see from the finite
temperature effective potential, which can be calculated from the zero
temperature classical potential in loop expansion.

\par In three spatial dimensions, four different kinds of topological
defects can arise. Whether or not topological defects will develop
during a symmetry breaking phase transition as well as the
determination of their type both depend on the topology of the vacuum
manifold ${\cal M}$. The properties of ${\cal M}$ are usually
described by the $n^{\rm th}$ homotopy group $\pi_n({\cal M})$ which
classifies dinstinct mappings from the $n$-dimensional sphere $S^n$
into the manifold ${\cal M}$.  If ${\cal M}$ has disconnected
components, or equivalently if the order $n$ of the non-trivial
homotopy group is $n=0$, then two-dimensional defects, called {\sl
domain walls}, form.  The spacetime dimension $d$ of the defects is
given in terms of the order of the non-trivial homotopy group by
$d=4-1-n$. If ${\cal M}$ is not simply connected, in other words if
${\cal M}$ contains loops which cannot be continuously shrunk into a
point, then {\sl cosmic strings} form. A necessary, but not
sufficient, condition for the existence of stable strings is that the
first homotopy group (the fundamental group) $\pi_1({\cal M})$ of
${\cal M}$, must be non-trivial, or multiply connected. Cosmic strings
are line-like defects, $d=2$. If ${\cal M}$ contains unshrinkable
surfaces, then {\sl monopoles} form, for which $n=2, ~d=1$. Finally,
if ${\cal M}$ contains non-contractible three-spheres, then {\sl
textures} form. Textures are event-like defects with $n=3, ~d=0$.

\par Topological defects are called local or global. The energy of
local defects is strongly confined, while the gradient energy of
global defects is spread out over the causal horizon at defect
formation.

\par In these lectures I will basically concentrate on cosmic strings,
which have very intriguing properties and interesting cosmological
implications~\cite{review-td,hk}. I will discuss the r\^ole which they
could play in the large-scale structure and the anisotropies of the
cosmic microwave background, as well as the r\^ ole of cosmic strings
in explaining various high energy phenomena.  I will then summarize
the gravitational effects of cosmic strings networks.  Finally, I will
discuss topological defects within brane world cosmology.

\par This review is organized as follows: In Section 2, I describe
cosmic strings in the context of field theory. I first discuss two
simple models, the abelian-Higgs model and the Goldstone model, which
lead to local and global strings respectively. I then summarize string
formation and string dynamics.  After this short introduction to the
physics of cosmic strings, I proceed with their physical/astrophysical
consequences.  In Section 3, I analyze the mechanism of cosmic
structure formation with topological defects. I first describe the
basic observable of the cosmic microwave background, and the physical
mechanisms which are responsible for the microwave background
anisotropies. I compare topological defects models versus inflationary
ones, as the two alternative mechanisms for structure formation within
gravitational instability theory. Concluding that topological defects
are excluded as the main source of the large-scale structure, I
address the question of finding the more {\sl natural} class of
inflationary models, meaning inflationary scenarios motivated by high
energy physics. I then discuss the implications of these findings for
particle physics models. In Section 4, I describe how cosmic strings
could explain various high energy phenomena (baryon asymmetry,
ultra-high energy cosmic rays, and gamma ray bursts). In Section 5, I
discuss the gravitational effects of cosmic strings networks.  In
Section 6, I briefly comment on the r\^ole of topological defects in
brane world cosmology. I close this review with the conclusions in
Section 7.

\section{Cosmic strings in field theory}
In two simple cases, I describe how cosmic strings appear in the
context of field theory. I first present the abelian-Higgs model and
I then proceed with the Goldstone model for which the gauge fields are
removed.  I will not discuss non-abelian strings (e.g., $Z_N-$
strings, Alice strings, electroweak strings) which can arise at a
symmetry breaking $G\rightarrow H$, with $G$ a non-abelian group, for
which the unbroken subgroup $H$ is disconnected.

\subsection{LOCAL OR GAUGE STRINGS}
Let us consider the simplest example of a gauge field theory with
SSB. This is the abelian-Higgs model with
Lagrangian density
\begin{equation}
{\cal L}=D_\mu{\bar \phi}D^\mu\phi -{1\over
4}F_{\mu\nu}F^{\mu\nu}-V(\phi) ~~,~~V(\phi)={\lambda\over
4}(|\phi|^2-\eta^2)^2~,
\label{ld2}
\end{equation}
where $D_\mu=\partial_\mu+ieA_\mu$ and $F_{\mu\nu}=\partial_\mu A_\nu
-\partial_\nu A_\mu$. The $U(1)$ invariance is satisfied by the
transformations $\phi\rightarrow \phi\exp[i\alpha(x)]$,
~$A_\mu\rightarrow A_\mu-(1/e)\partial_\mu \alpha(x)$, where
$\alpha(x)$ is a real single-valued function with a spatial
dependence.

\par The Euler-Lagrange equations, which follow from Eq.~(\ref{ld2}),
read
\begin{equation}
[D^2+\lambda(|\phi|^2-\eta^2)]\phi=0~~,~~
\partial_\nu F^{\mu\nu}+ie({\bar\phi}D^\mu\phi-D^\mu{\bar\phi}\phi)=0~.
\end{equation}
We are seeking a cylindrically symmetric, static solution of the above
field equations. In cylindrical polar cordinates $\{\rho,\varphi,z\}$,
we demand the solution to approach a vacuum state as $\rho\rightarrow
\infty$, meaning that $\phi\rightarrow\eta\exp(in\varphi)$ (with $n$
an integer) and $A_\mu\rightarrow (n/e)\partial_\mu\varphi$, thus
$D_\mu\phi\rightarrow 0$. We make the ansatz:
\begin{eqnarray}
\phi&=&\eta f(\rho)\exp(i n\varphi)\nonumber\\ A_x&=&{-n\eta\over
e}\left({y\over \rho^2}\right) \alpha(\rho)\nonumber\\
A_y&=&{n\eta\over e}\left({x\over \rho^2}\right)
\alpha(\rho)\nonumber\\ A_z&=&0~,
\end{eqnarray}
which we insert into the field equations and we  get two coupled
ordinary differential equations for $f$ and $\alpha$. Solutions to
this system, describing a string along the $z$-axis and satisfying the
requirements $f(0)=\alpha(0)=0$, and $f(\rho)~,~\alpha(\rho)
\rightarrow 1$ as $\rho\rightarrow 1$, can be found numerically.

\par The string solution of the abelian-Higgs model is the
Nielsen-Olesen vortex~\cite{nov}. For a local string, the energy per
unit length is finite.  For large $\rho$, the energy per unit length
of the gauge string, $\mu=\int\rho d\rho d\varphi T_0^0$, is of the
order of $\mu\sim 2\pi\eta^2$.  The local string also contains a tube
of magnetic flux, with quantized flux $(2\pi n/e)$.

\par The thickness of a Nielsen-Oleson string is about $1/\eta$ and on
length scales which are much larger than $1/\eta$, its energy momentum
tensor can be approximated by
\begin{equation}
(T_\mu^\nu)=\mu\delta(x)\delta(y){\rm diag}(1,0,0,1)~.
\end{equation}

\par The vortices in the abelian-Higgs model have as condensed matter
analogues, the flux tubes in superconductors.

\subsection{GLOBAL STRINGS}
Let us consider a complex scalar field $\phi(x)$, described by the
Lagrangian density
\begin{equation}
{\cal L}={1\over 2}\partial_\mu\phi\partial^\mu\bar{\phi}-V(\phi)~~,
~~V(\phi)={\lambda\over 4}(|\phi|^2-\eta^2)^2~.
\label{ld1}
\end{equation}
We remain at low temperatures, so we only consider the zero
temperature effective potential, while we do not include any thermal
corrections.  The Lagrangian density, Eq.~(\ref{ld1}), has a global
$U(1)$ symmetry, under the transformation $\phi\rightarrow \phi
\exp(i\tilde{\alpha})$, with $\tilde{\alpha}$ constant.

\par The Euler-Lagrange equations that follow from Eq.~(\ref{ld1})
read
\begin{equation}
[\partial^2+\lambda(|\phi|^2-\eta^2)]\phi=0~.
\end{equation}
Thus, the vacuum manifold is a circle of radius $\eta$.  At high
enough temperatures, $T\gg \eta$, the effective potential has a single
minimum located at $\phi=0$. However, as the temperature falls below a
critical value $T_{\rm c}$, with $T_{\rm c}\sim\eta$, a phase
transition occurs and the field $\phi$ takes a non-zero vacuum
expectation value, $\langle 0|\phi|0\rangle =\eta\exp(i\alpha_0)$,
where $\alpha_0$ is a constant.  This ground state solution is stable,
and since it does not remain invariant under the $U(1)$ symmetry
transformations, the symmetry is said to be broken by the vacuum.

\par Besides the vacuum, there are also static solutions with non-zero
energy density. In  cylindrical polar
coordinates $\{\rho,\varphi, z\}$, we make the ansatz:
\begin{equation}
\phi({\bf x})=\eta \tilde{f}(\rho)~\exp(in\varphi)
\end{equation}
(with $n$ an integer). The field equations of motion reduce to a
single non-linear ordinary differential equation for $\tilde{f}$.
As $\rho\rightarrow 0$, continuity of
$\phi$ demands $\tilde{f}\rightarrow 0$, while at infinity, $\tilde{f}
\rightarrow 1$, so that $\phi\rightarrow |\phi|=\eta$.  These
solutions are known as {\sl global strings} or {\sl vortices}
~\cite{ve82,vor}, and they have close analogies with the vortices in
superfluid ${}^4$He~\cite{hel1,hel2}.

\par The non-zero components of the vortex energy momentum tensor are
\begin{equation}
T_0^0=T_z^z=|\dot\phi^2|+V(\phi)+|{\bf\nabla}\phi|^2=-{\lambda\eta^4\over
2}\left[\tilde{f}^{'2}-{1\over 2}(\tilde{f}^2-1)^2+{\eta^2\over
\lambda\eta^2\rho^2}\tilde{f}^2\right]
\label{ed}
\end{equation}
The energy per unit length of cross-section of a vortex of radius $R$
is
\begin{equation}
\mu(R)=2\pi\int_0^R~T_0^0~\rho~{\rm
d}\rho~\sim~\pi\eta^2\ln(\sqrt{\lambda}\eta R)~.
\end{equation}
The logarithmic divergence for large $R\gg (\sqrt{\lambda}\eta)^{-1}$,
comes from the angular part of the gradient term, the last term in
Eq.~(\ref{ed}), which decays only as $1/\rho^2$. Clearly an upper
cutoff is naturally provided by either the curvature radius of the
string or by its distance to the next vortex.  

\par The energy per unit length of local strings was found to be
finite, since the angular derivative of the field was replaced by a
covariant derivative, which can vanish faster than $1/\rho$ at
infinity. The energy density of local strings was found to be much
more localized than in the global string case.

\par I will now discuss the formation and dynamics of cosmic strings,
making no distinction between local/global strings.

\subsection{FORMATION OF STRINGS}
\par According to the laws of standard cosmology, our universe in the
past was denser and hotter. Thus, the early universe was in a
symmetric phase and there were no topological defects. As the universe
expands, it cools down and once its temperature falls below a critical
value $T_{\rm c}$, the Higgs field settles down into the valley of the
minima of the potential. Eventually, as the temperature drops to zero,
the system tends towards one of the equivalent vacuum states and the
original symmetry is broken.

\par The density of defects depends on the order of the phase
transition. In the original picture~\cite{kibble76}, defects form at a
second-order phase transition and their density is determined by the
correlation length $\xi(T_{\rm f})$ of the Higgs field $\phi$ at the
{\sl freeze-out} temperature $T_{\rm f}$, which is identified with the
Ginzburg temperature $T_{\rm G}$.  The correlation length $\xi(T)$
sets the maximum distance over which the Higgs field can be correlated
during a symmetry breaking phase transition.  The Ginzburg temperature
$T_{\rm G}$ is the temperature below which thermal fluctuations cannot
restore the broken symmetry at the scale of the correlation
length. Since the horizon distance is finite, at the time of phase
transition the Higgs field must be uncorrelated on scales greater than
the particle horizon, which sets an absolute maximum for the
correlation length.  As a result, non-trivial vacuum configurations
will necessarily be produced during a symmetry breaking phase
transition in the early universe, with an abundance of order one per
horizon volume.  This original picture has been later revised, as I
will now discuss.

\par For given temperature $T$, the time $\tau(T)$ it takes to
correlations to establish on the length scale $\xi(T)$, has a rather
crucial r\^ ole~\cite{zurek1,zurek2}. As $T\rightarrow T_{\rm c}$,
both $\xi(T)$ and $\tau(T)$ diverge as inverse powers of $(T-T_{\rm
c})$. Once $\tau(T)$ exceeds the dynamical time scale $t_{\rm
D}=|T-T_{\rm c}|/|\dot T|$ of the temperature variation, then
correlations cannot establish on larger scales. Thus, the freeze-out
temperature $T_{\rm f}$ should be determined by $\tau\sim t_{\rm D}$
and this is known as the Kibble-Zurek
picture~\cite{zurek1,zurek2,kibble}.  In this picture of defect
formation, the distribution of defects is determined by the
distribution of zeros of the Higgs field $\phi$, smoothed over the
correlation length $\xi$, in thermal equilibrium at the freeze-out
temperature $T_{\rm f}$. This picture is supported by experiments in
$^3He$, by non-equilibrium quantum field theory studies, as well as by
numerical simulations. However, despite all this, defect formation in
second-order phase transitions still remains an open issue.

\par On the other hand, if defects were formed as a result of a
first-order phase transition, then their density depends on how
efficient is phase equilibration in bubble collisions. It turns out
that the physics of defect formation in bubble collissions is quite
involved and one cannot just use a simple rule in order to estimate
the density of defects. I will not enter into any details and I refer
the reader to the literature~\cite{1pt1,1pt2,1pt3,1pt4,1pt5}.

\par In the case of cosmic strings, energetics do not determine the
phase of the vacuum expectation value of the complex Higgs field
$\phi$, since the vacuum energy depends only upon $|\phi|$, while the
phase is arbitrary. Since $\phi$ must be single valued, the total
change in phase around any closed path must be an integer multiple of
$2\pi$. Imagine a closed path with total change in phase equal to
$2\pi$. As the path is shrunk to a point, the total change in phase
cannot change continuously from $2\pi$ to $0$. Thus, there must be one
point inside the path where the phase is undefined, meaning that the
vacuum expectation value of $\phi$ vanishes. The region of false
vacuum inside this path is part of a tube of false vacuum, the cosmic
string.  Cosmic strings must be either infinite in length or closed in
form of loops, since otherwise it would be possible to deform the path
around the tube and contract it to a point without encountering the
tube of false vacuum.  Assuming that the radius of curvature of a
string is much greater than its thickness, cosmic strings can be
considered as one-dimensional, in space, objects.

\subsection{STRING DYNAMICS}
The world history of a string can be expressed by a two-dimensional surface
in the four-dimensional spacetime, which is called the string worldsheet:
\begin{equation}
x^\mu=x^\mu(\zeta^a)~~,~~a=0,1
\end{equation}
where the worldsheet coordinates $\zeta^0, \zeta^1$ are arbitrary
parameters chosen so that $\zeta^0$ is timelike and $\zeta^1$
spacelike. 

\par The string equations of motion, in the limit of strings of zero
thickness, are derived from the Nambu-Goto action which, up to an
overall factor, corresponds to the surface area swept out by the
string in spacetime:
\begin{equation}
S=-\mu\int\sqrt{-\gamma}d^2\zeta~,
\label{nga}
\end{equation}
where $\gamma$ is the determinant of the two-dimensional worldheet
metric $\gamma_{ab}$,
\begin{equation}
\gamma={\rm det}(\gamma_{ab})={1\over
2}\epsilon^{ac}\epsilon^{bd}\gamma_{ab}\gamma_{cd}~~ ,
~~\gamma_{ab}=g_{\mu\nu}x^\mu_{,a}x^\nu_{,b}
\end{equation}
with $g_{\mu\nu}$ the four-dimensional metric.

\par Varying the Nambu-Goto action with respect to $x^\mu(\zeta^a)$ we
obtain the string equations of motion
\begin{equation}
x^{\mu~;a}_{,a}+\Gamma^\mu_{\nu\sigma}\Gamma^{ab}x^\nu_{,a}x^\sigma_{,b}=0~,
\end{equation}
with $\Gamma^\mu_{\nu\sigma}$ the four-dimensional Christoffel symbol.

\par The string energy-momentum tensor can be found by varying the
Nambu-Goto action with respect to the metric $g_{\mu\nu}$:
\begin{equation}
T^{\mu\nu}\sqrt{-g}=\mu\int d^2\zeta\sqrt{-\gamma}\gamma^{ab}x^\mu_{,a}
x^\nu_{,b}\delta^{(4)}(x^\sigma-x^\sigma(\zeta^a))~.
\end{equation}

\par One then distinguishes between strings in flat or in curved
spacetime, fix the gauge conditions, find the string equations of
motion, and solve them.

\par The Nambu-Goto action, Eq.~(\ref{nga}), describes the string
motion in the absence of string intersections with itself or another
string. At least within simple models, strings cannot simply pass
through one another with no interaction, so strings reconnect. At
every intercommuting, string pieces moving originaly independently
from each other, become connected. At the intersection time, ${\bf
x}'$ and $\dot {\bf x}$ change very rapidly as functions of $\zeta^1$
on a length scale which is comparable to the thickness of the
string. The region of this rapid change is known as {\sl kink}, and at
the kink, ${\bf x}'$ and $\dot{\bf x}$ are discontinuous. Closed loops
of string can self-intersect and break into smaller loops.

\par For many years, the evolution of a string network was described
in the basis of the {\sl one-scale model}~\cite{kibble}, which gives
long-string evolution in terms of a correlation length , given two
free parameters, the loop chopping efficiency and the root-mean-square
velocity.  However, strings in an interacting network develop
substantial substructure in the form of kinks and wiggles, on a
smaller scale than the characteristic length scale of the
network. Numerical studies~\cite{sa,sv} have revealed the wiggliness
of the long strings and have demonstrated the inedequacy of the
one-scale model. As a result, the {\sl three-scale model} has been
proposed~\cite{ack}.  A numerical study of the three-scale model
emphasizing the important dependence of small-scale structure on the
numerical loop cutoff has been presented in Ref.\cite{vhs}.

\par In the next sections I will discuss the possible r\^ ole of
cosmic strings on various cosmological and astrophysical issues.

\section{Cosmic structure formation with topological defects}
The geometry of our universe is to a very good approximation
isotropic, and assuming that we are not in a special position, then it
is also homogeneous. The best observational evidence in support to
this is the isotropy of the cosmic microwave background (CMB).  The
CMB, last scattered at the epoch of decoupling, has to a high accuracy
a black-body distribution~\cite{bb,cheng}, with a temperature
$T_0=2.728\pm 0.002 ~{\rm K}$, which is almost independent of
direction.  The DMR experiment on the COBE satellite measured a tiny
variation $\Delta T$ in intensity of the CMB, at fixed frequency. This
is equivalently expressed as a variation in the temperature, which was
measured~\cite{cobe1,cobe2} to be $\Delta T/T_0 \approx 10^{-5}$. On
the other hand, on galaxy and cluster scales the matter distribution
is very inhomogeneous.

\par The origin of the large-scale structure in the universe, remains
one of the most important questions in cosmology.  Within the
framework of gravitational instability, two families of models
attempted to explain the formation of the structure one observes.
Initial density perturbations can either be due to {\sl freezing in}
of quantum fluctuations of a scalar field during an inflationary
period, or they may be seeded by a class of topological defects, which
could have formed naturally during a symmetry breaking phase
transition in the early universe.  The CMB anisotropies provide a link
between theoretical predictions and observational data, which may
allow us to distinguish between inflationary models and topological
defects scenarios, by purely linear analysis.  More precisely, the
characteristics of the CMB anisotropy multipole moments (position,
amplitude of acoustic peaks), and the statistical properties of the
CMB anisotropy are used to discriminate among models, and to
constrain the parameters space~\cite{mspain}.

\subsection{OBSERVABLES OF THE CMB}
The basic observable is the CMB intensity as a function of frequency
and direction of observation $\hat{{\bf n}}$.  We want to calculate
temperature anisotropies in the sky, thus it is natural to expand
$\Delta T/T_0$ in spherical harmonics:
\begin{equation}
{\Delta T\over T_0} (\hat{{\bf n}})=\sum_{\ell=1}^{\infty}
\sum_{m=-\ell}^{m=\ell}a_{\ell m}Y_{\ell m}(\hat{{\bf n}}){\cal W}_\ell~,
\end{equation}
where $\Delta T=T-T_0$, with $T_0$ the mean temperature in the sky,
and ${\cal W}_\ell$ stands for the $\ell$-dependent window function of
the particular experiment.

\par The angular power spectrum of CMB anisotropies is expressed in
terms of the dimensionless coefficients $C_\ell$, which is the
ensemble average of the coefficients $a_{\ell m}$:
\begin{equation}
C_\ell = \langle |a_{\ell m}|^2\rangle~.
\end{equation}
If the fluctuations are statistically isotropic, then $C_\ell$'s are
independent of $m$, and if they are Gaussian, then all the statistical
information is contained in the power spectrum.  Topological defects
generate non-Gaussian perturbations, while one-field inflation leads,
in general, to Gaussian perturbations.  Departure from the vacuum
initial conditions for cosmological perturbations of quantum
mechanical origin leads generically to a non-Gaussian signature,
however the signal-to-noise ratio is far away from experimental
detection~\cite{mrs,gms}.  In what follows when I refer to
perturbations of quantum-mechanical origin, I will simply consider as
initial state to be the vacuum.

\par The relation between the power spectrum and the two-point correlation
function is:
\begin{equation}
\left\langle{\Delta T\over T_0}(\hat{{\bf n}}_1)
{\Delta T\over T_0}(\hat{{\bf n}}_2)\right\rangle =
{1\over 4\pi}\sum_\ell (2\ell+1)C_\ell P_\ell(\hat{{\bf n}}_1\cdot
\hat{{\bf n}}_2)~,
\end{equation}
where $P_\ell$ denotes the Legendre polynomials. The angular power
spectrum compares points in the sky separated by an angle $\vartheta$,
where $(\hat{{\bf n}}_1\cdot \hat{{\bf n}}_2)=\cos\vartheta$.  Here
the brackets denote spatial average, or expectation values if
perturbations are quantized.  The value of $C_\ell$ is determined by
fluctuations on angular scales of order $2\pi/\ell$. 

\par In a real experiment, we have only one universe and one sky, and
therefore we cannot measure an ensemble average. Assuming statistical
isotropy,
\begin{equation}
C_\ell\simeq C_\ell^{\rm obs}={1\over 2\ell+1}\sum_m |a_{\ell m}|^2~,
\end{equation}
which in the ideal case of full sky coverage, this yields an average
on $2\ell+1$ numbers.  Assuming that the temperature fluctuations are
Gaussian, the observed mean deviates from the ensemble average by
about
\begin{equation}
{\sqrt{(C_\ell^{\rm obs})^2-(C_\ell)^2}\over C_\ell}
\simeq \sqrt{{2\over 2\ell+1}}~.
\end{equation}
This limitation of the precision of a measurement is known as {\sl
cosmic variance} and it is important especially for low multipoles.

\subsection{PHYSICS OF THE CMB}
Since the distribution of photons is uniform, perturbations are small,
and they can be studied with linear perturbation theory.  One can
split perturbations into scalar, vector and tensor modes; different
components do not mix.  Initially vector perturbations rapidly decay,
while scalar and tensor perturbations contribute to the CMB.  At the time
of recombination ($T_{\rm rec}\sim 3000 K$, $z_{\rm rec}\approx
10^3)$, electrons and protons formed neutral hydrogen.  Earlier, free
electrons acted as glue between the photons and the baryons through
Thomson and Coulomb scattering, implying that the cosmological plasma
was a tightly coupled photon-baryon fluid. After recombination, the
universe becomes transparent for CMB photons, and they move along
geodesics of the perturbed Friedmann geometry.

\par Integrating the perturbed geodesic equation, temperature
anisotropies in {\sl gauge-invariant form} take the
form~\cite{sasaki,RuthReview,rd}:
\begin{eqnarray}
\left({\Delta T\over T}\right)^{({\rm s})}(\eta_0,{\bf x}_0,
{\hat{\bf n}})&=&{1\over 4}D_{\rm r}(\eta_{\rm dec},{\bf x}_{\rm
dec})+v_i(\eta_{\rm dec},{\bf x}_{\rm dec})n^i\nonumber\\
&&+(\Phi-\Psi)(\eta_{\rm dec},{\bf x}_{\rm dec})
-\int_{\eta_{\rm dec}}^{\eta_0}(\dot\Phi-\dot\Psi)(\eta,{\bf
x}(\eta)){\rm d}\eta\nonumber\\
\left({\Delta T\over T}\right)^{({\rm
t})}(\eta_0,{\bf x}_0, {\hat{\bf n}})&=&- \int_{\eta_{\rm
dec}}^{\eta_0} \dot h_{i j} (\eta,{\bf x}(\eta)) n^i n^j {\rm d}\eta~,
\label{ta}
\end{eqnarray}
where $\eta$ denotes conformal time; ${\bf x}(\eta)$ is the comoving
unperturbed photon position at time $\eta$; $D_{\rm r}$ is the photon
energy density fluctuations; $v_i$ is the baryon velocity field;
$\Phi, \Psi$ are the Bardeen potentials, an overdot denotes derivative
with respect to conformal time $\eta$; and $h_{i j}$ is a spatial
metric perturbation (tracelless $\delta^{i j} h_{i j}=0$ and
transverse $\vartheta_i h_{i j}=0$).

\par Let us take the scalar mode of temperature anisotropies, which is
the first of Eqs.~(\ref{ta}).  The first term describes the intrinsic
inhomogeneities on the surface of the last scattering due to acoustic
oscillations prior to decoupling. It also contains contributions to
the geometrical perturbations. The second term describes the relative
motions of emitter and observer. This is the Doppler contribution to
the CMB anisotropies. It appears on the same angular scale as the
acoustic term and we denote the sum of the acoustic and Doppler
contributions by {\sl acoustic peaks}. The last two terms are due to
the inhomogeneities in the space-time geometry; the first contribution
determines the change in the photon energy due to the difference of
the gravitational potential at the position of emitter and
observer. Together with the part contained in $D_{\rm r}$ they
represent the {\sl ordinary} Sachs-Wolfe effect. The second term
accounts for red-shifting or blue-shifting caused by the time
dependence of the gravitational field along the path of the photon
({\sl Integrated Sachs-Wolfe} (ISW) effect). The sum of the two terms
is the full Sachs-Wolfe contribution (SW).

\par On angular scales $0.1^\circ\stackrel{<}{\sim}
\theta\stackrel{<}{\sim} 2^\circ$, the main contribution to the CMB
anisotropies comes from the acoustic peaks, while the SW effect is
dominant on large angular scales. For topological defects models, the
gravitational contribution is mainly due to the ISW; the ordinary
Sachs-Wolfe term even has the wrong spectrum, a white noise spectrum
instead of a Harrison-Zel'dovich~\cite{ruthzhou} spectrum.

\par On scales smaller than about $0.1^\circ$, the anisotropies are damped due
to the finite thickness of the recombination shell, as well as by
photon diffusion during recombination (Silk damping).  Baryons and
photons are very tightly coupled before recombination and oscillate as
one component fluid.  During the process of decoupling, photons slowly
diffuse out of over-dense into under-dense regions. To fully account
for this process, one has to solve the Boltzmann equation.

\subsection{Topological defects models versus inflationary ones}
Inflationary fluctuations are produced at a very early stage of the
evolution of the universe, and are driven far beyond the Hubble radius
by inflationary expansion. Subsequently, they are not altered anymore
and evolve freely according to homogeneous linear perturbation
equations until late times. These fluctuations are termed {\sl
passive} and {\sl coherent}. Passive, since no new perturbations are
created after inflation; coherent since randomness only enters the
creation of perturbations during inflation, subsequently they evolve
in a deterministic and coherent manner.  

\par Within inflation, the induced perturbations are adiabatic density
perturbations, meaning that the density of each particle species is a
unique function of the total energy density $(\delta\rho_{\rm
m}/\rho_{\rm m}) =(3/4)(\delta\rho_{\rm r}/\rho_{\rm r})$.

\par The main difference in linear cosmological perturbation theory
for models with topological defects, as compared to the adiabatic
inflationary case, is that perturbations are generated by {\sl seeds}
(sources).  The seeds are defined as any non-uniformly distributed
form of energy, which contributes only a small fraction to the total
energy density of the universe and which interacts with the cosmic
fluid only gravitationally. Topological defects are a concrete example
of such seeds. The energy momentum tensor of the seeds enters in the
perturbation equation as a source term on the rhs, while the seeds
themselves evolve according to the background space-time;
perturbations in the seed evolution are of second order.

\par Topological defects (and in general models with {\sl seeds}) lead
to isocurvature density perturbations, in the sense that the total
density perturbation vanishes, but those of the individual particle
species do not. In other words, there is a non-zero temperature
perturbation $S=(\delta\rho_{\rm m}/ \rho_{\rm m})-(3/ 4)
(\delta\rho_{\rm r}/ \rho_{\rm r})\neq 0$,
To have isocurvature perturbations, the universe has to possess more
than the single degree of freedom provided by the total energy
density.  For isocurvature perturbations, the measure of the spatial
curvature seen by comoving observers is zero, ${\cal R}=0$, during
RDE, but on large scales entering the horizon during MDE, ${\cal
R}=(1/3)S$.

 \par In models with topological defects,
fluctuations are generated continuously and evolve according to
inhomogeneous linear perturbation equations.  The energy momentum
tensor of defects is determined by the their evolution which, in
general, is a non-linear process. These perturbations are called {\sl
active} and {\sl incoherent}. Active since new fluid perturbations are
induced continuously due to the presence of the defects; incoherent
since the randomness of the non-linear seed evolution which sources
the perturbations can destroy the coherence of fluctuations in the
cosmic fluid.  The highly non-linear structure of the topological
defects dynamics makes the study of the evolution of these causal
(there are no correlations on super-Horizon scales) and incoherent
initial perturbations much more complicated.

\par Among the various topological defects, local cosmic strings or
any kind of global defects could, in principle, induce the initial
perturbations. Models with domain walls or local monopoles are ruled
out. In the first case, because even a single domain wall stretching
across the present universe would overclose it, while in the second
one, since local monopoles soon dominate the energy density of the
universe due to the lack of interactions. Finally, textures cannot
exist as local coherent defects which have a non-vanishing gradient
energy.  On the other hand, as it was shown from numerical
simulations~\cite{sv}, a network of local cosmic strings approaches a
scaling distribution, meaning that once scaling has been reached,
there only be a few long strings crossing each Hubble volume, together
with a distribution of rather small closed loops of strings. Thus,
cosmic strings are not cosmologically undesirable. Global defects are
also acceptable as possible candidates for a scenario of structure
formation, since there are long range forces between them resulting to
a scaling solution, in the sense that there is again a fixed number of
such defects per Hubble volume.

\par The measurements of cosmic microwave background anisotropy by the
COBE-DMR experiment provide the normalization $T_{\rm c}^2/M_{\rm Pl}
\sim 10^{-5}$, where $T_c$ denotes the temperature at the symmetry
breaking phase transition which gave rise to the topological defects,
within a scenario of structure formation via a mechanism of
seeds. Thus, to seed the observed large scale structure, global
defects or local cosmic strings should have been formed at $ T_{\rm
c}\sim 10^{16}$ GeV, which is the scale of unification of weak, strong
and electromagnetic interactions. 

\par Within linear cosmological perturbation theory, structure
formation induced by seeds is determined by the solution of the
inhomogeneous equation
\begin{equation} 
\DD X({\bf k},t) = \SS({\bf k},t)~, 
\end{equation}
where $X$ is a vector containing all the background perturbation
variables for a given mode specified by the wave-vector ${\bf k}$,
like the $a_{lm}$'s of the CMB anisotropies, the dark matter density
fluctuation, the peculiar velocity potential etc., $\DD$ is a linear
time-dependent ordinary differential operator, and the source term $
\SS$ is given by linear combinations of the energy momentum tensor of
the seed (the type of topological defects we are considering).  The
generic solution of this equation is given in terms of a Green's
function and has the following form~\cite{vest}
\be 
X_i({\bf k},t_0) = \int_{t_{in}}^{t_0}\GG_{il}({\bf k},t_{0},t)
\SS_l({\bf k}, t)dt~.  
\ee 
At the end, we need to determine expectation values, which are given
by 
\be 
\langle X_i(\bk ,t_0)X_j(\bk,t_0)^*\rangle
=\int_{t_{in}}^{t_0}\int_{\eta_{in}}^{\eta_0}
\GG_{il}(t_{0},t)\GG_{jm}^*(t_{0},t') \langle
\SS_l(t)\SS_m^*(t')\rangle dt dt' .
\label{pow}
\ee Thus, the only information we need from topological defects
simulations in order to determine cosmic microwave background and
large-scale structure power spectra, is the {\sl unequal time
two-point correlators}~\cite{cor}, $\langle
\SS_l(t)\SS_m^*(t')\rangle$, of the seed energy-momentum tensor. This
problem can, in general, be solved by an eigenvector expansion
method~\cite{pen}.  \par A crucial issue, is how to
reach the desirable dynamical range in order to compute accurately
observables which will then be compared to the data. To overcome this
difficulty, one applies the theoretical requirements of causality,
scaling, and statistical homogeneity and isotropy.

\par The complexity of computations in defect scenarios has of course
affected the number of studies, in which the mechanism of structure
formation is provided by seeds, as compared to the inflationary case.
Nevertheless, one can find in the literature scenario with global
defects or local cosmic strings, as well as models with causal seeds
which attempt to imitate inflation. Let us summarize the present
status, once we compare theoretical predictions with the available
data.

\par On large angular scales ($\ell \laq 50$), defect models lead to
the same prediction as inflation, namely, they both predict an
approximately scale-invariant (Harrison-Zel'dovich) spectrum of
perturbations. Their only difference concerns the statistics of the
induced fluctuations. Inflation predicts generically Gaussian
fluctuations, whereas in the case of topological defects models, even
if initially the defect energy-momentum tensor would be Gaussian,
non-Gaussianities will be induced from the non-linear defect
evolution. Thus, in defect scenarios, the induced fluctuations are
non-Gaussian, at least at sufficiently high angular resolution.  This
is an interesting fingerprint, even though difficult to test through
the data.

\par On itermediate and small angular scales however, the predictions
of models with seeds are quite different than those of inflation, due
to the different nature of the induced perturbations.  In topological
defects models, defect fluctuations are constantly generated by the seed
evolution.  The non-linear defect evolution and the fact that the
random initial conditions of the source term in the perturbation
equations of a given scale leak into other scales, destroy perfect
coherence.  The incoherent aspect of active perturbations does not
influence the position of the acoustic peaks, but it does affect the
structure of secondary oscillations, namely secondary oscillations may
get washed out. Thus, in topological defects models, incoherent
fluctuations lead to a single bump at smaller angular scales (larger
$\ell$), than those predicted within any inflationary scenario.  This
incoherent feature is shared in common by local and global defects.

\par Studying the acoustic peaks for perturbations induced by global
textures, we found~\cite{ram} that the amplitude of the first acoustic
peak is only $\sim 1.5$ times higher than the SW plateau and its
position is at $\ell\sim 350$.  Global ${\cal {O}}(N)$ textures in the
large $N$ limit show a flat spectrum, with a slow decay after
$\ell\sim 100$~\cite{rma}.  There are similar results with other
global ${\cal {O}}(N)$ defects~\cite{neil98,pen}. For local cosmic
strings the predictions range from an almost flat spectrum to a single
wide bump at $\ell\sim 500$ with extremely rapidly decaying
tail~\cite{cs1,cs2,cs3,cs4,cs5}.  This discrepancy between global and
local defects may come from the larger difference between the
energy-momentum tensor in the radiation and matter-dominated era of
the local defects, as compared to the case of global defects.
Moreover, it seems that the microphysics of the cosmic string network
plays a crucial role in the height and in the position of the bump. It
is interesting to mention that in the case of local cosmic strings,
the energy density is clearly dominant over the other components of
the energy-momentum tensor, whereas this feature is not shared by
global defects. In addition, the angular cosmic microwave background
power spectrum of local cosmic strings depends on the equation of
state of the decay product.

\par One expects also to find a defect fingerprint on the power
spectrum of polarization. Due to the presence of vector perturbations,
all defect models predict a much larger component of magnetic-type
polarization on small angular scales, as compared to a standard cold
dark matter model. In addition, the isocurvature shift of the acoustic
peaks, as well as the incoherent character of the perturbations,
appear in the polarization signal.

\par Standard inflation predicts the position of
the first peak at $\ell\sim 220$ and its amplitude $\sim(4-6)$ times
higher than the SW plateau.

\par One can manufacture models~\cite{ruthmairi} with structure
formation being induced by {\sl scaling seeds}, which lead to an
angular power spectrum with the same characteristics (position and
amplitude of acoustic peaks), as the one predicted by standard
inflationary models.  The open question is, though, whether such
models are the outcome of a realistic theory.  At this point, I would
like to remind to the reader that the question whether or not
inflationary models which fit the data are physical or not, has also
to be addressed, even though people often tend to {\sl forget} about it.

\par I would also like to briefly mention another model, which could
account for the origin of the large-scale curvature perturbation in
our universe. This is the so-called {\sl curvaton
model}~\cite{lw2002}, where the curvaton is a scalar field which
remains light during inflation, and gets perturbed with an almost
scale-invariant spectrum. Initially there is an isocurvature density
perturbation, which generates the curvature perturbation later, when
the curvaton density becomes a significant fraction of the total. A
priori this model can be applied in scenarios with seeds.

\par The position and amplitude of the acoustic peaks, as found by the
CMB measurements --- and in particular by the BOOMERanG~\cite{boom,boom1},
MAXIMA~\cite{max,max1}, and DASI~\cite{dasi,dasi1} experiments --- are in
disagreement with the predictions of topological defects models.
Thus, the latest CMB anisotropy measurements rule out pure topological
defects models as the origin of initial density fluctuations.

\par The inflationary paradigm is at present the most appealing
candidate for describing the early universe. However, inflation is not
free from open questions and I would like to mention the following
three types of issues, to which any inflationary model should give an
answer:\\ (i)It is difficult to implement inflation in high energy
physics. More precisely, the inflaton potential coupling constant must
be very low in order to reproduce the CMB data. This is related to the
question of deciding which kind of inflationary model is the more
natural one.\\ (ii) The quantum fluctuations are typically generated
from sub-Planckian scales and therefore one should examine the
validity of the theoretical predictions based upon standard quantum
mechanics. Recent studies~\cite{jetal} seem to indicate that inflation
is robust to some changes of the standard laws of physics beyond the
Planck scale. \\ (iii) It is almost always assumed that the initial
state of the perturbations is the vacuum. The proof of such a
hypothesis, if it exists, should rely on full quantum gravity, a
theory which is still lacking. If the initial state is not the vacuum,
this would imply a large energy density of inflaton field quanta, not
of a cosmological term type~\cite{ll}. Thus, non-vacuum initial states
lead to a back-reaction problem, which has not been calculated yet.

\par Since inflation provides, at present, the most appealing
candidate for describing the early universe, we face the choice of the
more physical inflationary scenario. Following the philosophy that the
more natural cosmological model is the one which arises from particle
physics models, as for example superstring theories, we will see that
topological defects, and in particular cosmic strings, can still play
a r\^ ole for the origin of the large-scale structure and the patterns
of cosmic microwave sky.  I will thus comment on a new degeneracy
apparently arising in the CMB data, that would be due to a small ---
but still significant --- contribution of topological defects.

\par In many particle physics based models, inflation ends with the
formation of topological defects, and in particular cosmic
strings~\cite{linde2,linde3,lr}.  Moreover, cosmic strings are
predicted by many realistic particle physics models.  Thus, even
though the current CMB anisotropy measurements seem to rule out the
class of generic topological defects models as the unique mechanism
responsible for the CMB fluctuations, it is conceivable to consider a
mixed perturbation model, in which the primordial fluctuations are
induced by inflation with a non-negligible topological defects
contribution.  

\par We consider~\cite{bprs} a model in which a network of cosmic
strings evolved independently of any pre-existing fluctuation
background, generated by a standard cold dark matter with a non-zero
cosmological constant ($\Lambda$CDM) inflationary phase. As we shall
restrict our attention to the angular spectrum, we are in the
linear regime.  Thus,
\begin{equation}
C_\ell =   \alpha     C^{\scriptscriptstyle{\rm I}}_\ell
         + (1-\alpha) C^{\scriptscriptstyle{\rm S}}_\ell~,
\label{cl}
\end{equation}
where $C^{\scriptscriptstyle{\rm I}}_\ell$ and $C^{\scriptscriptstyle
{\rm S}}_\ell$ denote the (COBE normalized) Legendre coefficients due
to adiabatic inflation fluctuations and those stemming from the string
network respectively. The coefficient $\alpha$ in Eq.~(\ref{cl}) is a
free parameter giving the relative amplitude for the two
contributions.  We have to compare the $C_\ell$, given by
Eq.~(\ref{cl}), with data obtained from CMB anisotropy
measurements. 
\begin{figure}
\centering 
\vskip.5truecm
\epsfig{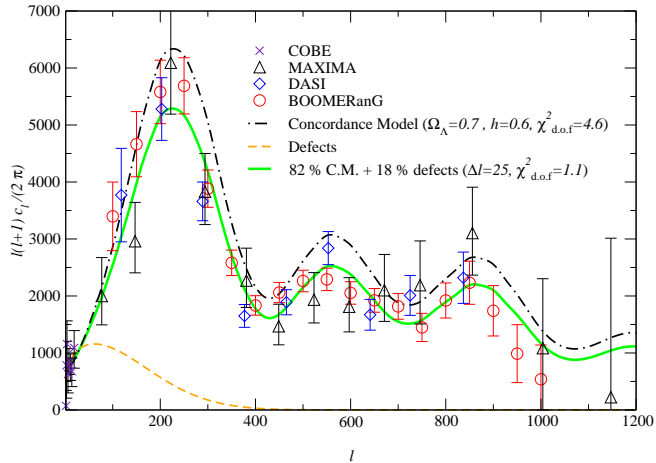}
\caption{$\ell (\ell + 1) C_\ell$ versus $\ell$ for three different
models. The upper dot-dashed line represents the prediction of a
$\Lambda$CDM model, with $n_{\rm S} = 1$, $\Omega_\Lambda = 0.7$,
$\Omega_{\rm m} = 0.3$, $\Omega_{\rm b} = 0.05$ and $h = 0.6$.  The
lower dashed line is a typical string spectrum. Both lines are
normalized at the COBE scale (crosses).  Combining both curves with
the parameter $\alpha$ produces the solid curve, with a $\chi^2$ per
degree of freedom slightly above unity. The string contribution is
about 18\% of the total.}
\label{fig1}
\end{figure}
Figure~\ref{fig1} shows the two uncorrelated spectra ($\Lambda$CDM
model, strings) as a function of $\ell$, both normalized on the COBE
data, together with the weighted sum. One clearly sees that neither
the upper dot-dashed line --- which represents the prediction of a
$\Lambda$CDM model, with $n_{\rm S} = 1$, $\Omega_\Lambda = 0.7$,
$\Omega_{\rm m} = 0.3$, $\Omega_{\rm b} = 0.05$ and $h = 0.6$ --- nor
the lower dashed line --- which is a typical string spectrum --- fit
the BOOMERanG, MAXIMA and DASI data (circles, triangles and
diamonds respectively).  The best fit~\cite{bprs}, having $\alpha \sim
0.82$, yields a non-negligible string contribution, although the
inflation produced perturbations represent the dominant part for this
spectrum.

\par This mixed perturbation model leads to the following
conclusions~\cite{bprs}:\\ (i) It seems still a bit premature to rule
out any contribution of cosmic strings to the CMB anisotropy
measurements, even though we conclude that pure topological defects
models are excluded as the mechanism of structure formation.\\ (ii)
There is some degree of degeneracy between the class of models with a
string contribution and those without any strings but with more widely
accepted cosmological parameters. We thus suggest to add the string
contribution as a new parameter to the standard parameters space.

\par I will next discuss the implications of the CMB constraints on
particle physics models.

\subsection{IMPLICATIONS FOR PARTICLE PHYSICS MODELS}
Supersymmetry is a well accepted framework for constructing extensions
of the standard model. Supersymmetry (SUSY) is either a global
symmetry, or a local symmetry, in which case it also includes gravity
and is called supergravity (SUGRA).  We would like to consider
inflationary models predicted within supersymmetric theories,
following the philosophy that such models of inflation should be the
more natural ones.  In the context of spontaneous broken global SUSY,
the scalar potential is the sum of $F$-terms and $D$-terms, and
therefore inflation can come from the non-zero vacuum expectation
value of either an $F$-term or by that of a $D$-term, which comes from
a gauge group $U(1)$. In both cases, one usually gets {\sl hybrid}
inflation.

\par Let us consider models where inflation is due to the non-zero
vacuum expectation value of a $D$-term. It was shown that at the end
of hybrid inflation, the formation of stable cosmic strings may occur
in the context of global SUSY theories~\cite{rj96}, as well as in the
context of SUGRA theories~\cite{linde3}.  Cosmic strings formed at the
end of $D$-term inflation are very heavy and temperature anisotropies
may arise both from inflationary dynamics and from the presence of
cosmic strings.  In the simplest version of $D$-term inflation, cosmic
strings may contribute to the angular spectrum an amount of order of
$75\%$~\cite{rj97}.  In the light of the findings of Ref.~\cite{bprs},
such a high contribution is not allowed from the current CMB
data. Thus, CMB measurements either rule out the simplest version of
$D$-term inflation, or, at least, modify the allowed window for its
free parameters.

\par I believe that the crucial question of finding the class of
 natural inflationary models, remains unfortunately still open. Only
 once this question is satisfactorily answered, we can indeed say that
 {\sl inflation won}. In this sense, this is a beautiful example of
 the fruitful interplay between cosmology and high energy physics.

\par The CMB measurements can also impose constraints on decaying
defects.  More precisely, one can constrain models with decaying
topological defects, demanding that the photons produced during the
decay should not lead to spectral distortions of the CMB in
disagreement with the limits put from the COBE/FIRAS. The strongest
limits are within theories with decaying vortons.

\par Let me remind that cosmic strings can become superconducting if
electromagnetic gauge invariance is broken inside the strings. The
original idea that strings could become superconducting was first
suggested by Witten~\cite{witten85a}. The electromagnetic properties
of superconducting strings are very similar to those of thin
superconducting wires. Since the thickness of the string is typically
within the electromagnetic penetration depth, one concludes that
superconducting strings can be penetrated by electric and magnetic
fields. One can distinguish between bosonic and fermionic string
superconductivity.  Loops of superconducting cosmic strings stabilized
by a current are called vortons~\cite{vort1,vort2} and they arise in
symmetry breaking theories above the electroweak scale for which the
vacuum manifold is not simply connected. Vortons are either infinite
or closed loops which decay via emission of gravitational radiation.
They can be characterized by two integer numbers, a topological one,
which  specifies the winding number of the current carrier phase
around the loop, and a dynamically conserved number, which is related,
in the charge-coupled case, to the total amount of electric charge.

\par The constraint on the non-thermal fractional energy density
production in photons reads
\begin{equation}
(\delta\rho_\gamma)/\rho_\gamma\leq 7\times 10^{-5}~.
\end{equation}
Topological defects that decay into photons at temperature
$\Theta_{\rm d}$ corresponding to a redshift of less than $10^6$ will
produce spectral distortions of the CMB.  In the case of theories with
decaying vortons the COBE/FIRAS constraint becomes~\cite{betal02}:
\begin{equation}
\left({\Theta_{\rm Q}\over m_{\rm Pl}}\right)^{5/4} \left({\Theta_{\rm
Q}\over \Theta_{\rm c}}\right)^{3/2} {\Theta_{\rm Q}\over \Theta_{\rm
d}} < 7\times 10^{-5}~,
\end{equation}
where $\Theta_{\rm Q}$ stands for the time when the string becomes
current-carrying and $\Theta_{\rm c}$ denotes the time of string
formation.

\par Using the CMB data we can therefore constrain the various
parameters of a model which predicts decaying topological defects.

\section{Cosmic strings and high energy phenomena}
I proceed with a brief presentation of the possible r\^ ole cosmic
strings can play on a number of high energy phenomena. As such, I
discuss the issues of baryogenesis, ultra-high energy cosmic rays, and
gamma ray bursts.

\subsection{Cosmic strings and baryogenesis}
One of the most important issues in modern cosmology is to find a way
to explain the observed asymmetry between matter and antimatter
(baryon (B) asymmetry).  In particular, we would like to explain the
observed value of the net baryon-to-entropy ratio at the present time,
which is
\begin{equation}
{\Delta n_{\rm B}\over s}(t_0)\sim 10^{-10}~,
\end{equation}
(where $s$ denotes the entropy density) starting from initial
conditions in the very early universe when this ratio vanishes. As it
was pointed out by Sakharov~\cite{Sakharov 1967} three basic criteria
must be satisfied in order to have a chance to explain the data,
namely: (i) the theory describing the microphysics must contain
baryon number violating processes, (ii) these processes must be C and
CP violating, and (iii) these processes must occur out of thermal
equilibrium.

\par These, necessary but not sufficient, criteria can be satisfied in
GUT theories, where baryon number violating processes are mediated by
superheavy Higgs and gauge particles.  Let us however examine the
magnitude of the predicted $n_{\rm B}/s$. It depends on the asymmetry
$\epsilon$ per decay, on the coupling constant $\lambda$ of the
$n_{\rm B}$ violating processes, and on the ratio $n_{\rm X}/s$ of the
number density $n_{\rm X}$ of superheavy Higgs and gauge particles to
the number density of photons, evaluated at the time $t_{\rm d}$ when
the $n_{\rm B}$ violating processes fall out of thermal equilibrium
(after the phase transition). The parameter $\epsilon$ is propotional
to the CP-violation parameter in the model, and in a GUT theory the
CP-violation parameter can be large (of order 1).  However, the ratio
$n_{\rm B}/s$ depends also on the value of the ratio $n_{\rm X}/s$,
and to evaluate it we must consider two different cases. If $T_{\rm d}
> m_{\rm X}$ then $n_{\rm X}\sim s$, while if $T_{\rm d} < m_{\rm X}$
then $n_{\rm X}$ is diluted exponentially in the time interval between
the time when $T=m_{\rm X}$ and the time when $T=T_{\rm d}$. Thus, in
this second cacse, $n_{\rm B}/s$ is exponentially suppressed, $(n_{\rm
B}/s) \sim (\lambda^2/g)\epsilon e^{-m_{\rm X}/T_{\rm d}}$ (where $g$
is the number of spin degrees of freedom in thermal equilibrium at the
time of the phase transition), implying that the standard GUT
baryogenesis mechanism is ineffective.

\par In the case where the standard GUT baryogenesis mechanism is
inefficient, topological defects, which are indeed out of equilibrium
configurations, may be helpful. As we have already discussed,
topological defects will inevitably be produced in the symmetry
breaking GUT transition, with the GUT symmetry restored inside the
defects. In an analogous way as the decay of free $X$ quanta, the $B-$
and $CP-$ violating decays of superheavy particles emitted by defects
may produce the required baryon asymmetry of the universe.  The
particles can be emitted either from the cusps of the strings, or from
small loops, or even during the decay of {\sl hybrid defects} (e.g.,
walls bounded by strings, or monopoles connected by strings). In
addition, baryon asymmetry can be also produced by decaying vortons.

\par Let us briefly see how decay of topological defects can produce a
non-vanishing $n_{\rm B}/s$~\cite{bdh91}.  For $m_{\bf X}< T_{\rm d}$,
when the $n_{\rm B}$ violating processes fall out of equilibrium, the
energy density in free $X$ quanta is much larger than the defect
density, implying that the defect driven baryogenesis is subdominant.
However, if $m_{\bf X}> T_{\rm d}$, then the energy density in free
quanta decays exponentially, while, in contrast, the density in
defects only decreases as a power of time and it will therefore soon
dominate baryogenesis. More precisely, it was shown that~\cite{bdh91}
\begin{equation}
\left({n_{\rm B}\over s}\right)|_{\rm defect}\sim \lambda^2{T_{\rm
d}\over \eta} \left({n_{\rm B}\over s}\right) |_0~,
\end{equation}
where $({n_{\rm B}\over s})|_0$ denotes the unsuppressed value using
GUT baryogenesis mechanism, and $\eta$ is the symmetry breaking scale.

\par A baryon asymmetry generated in the early universe can
be erased by non-perturbative electroweak processes. Thus, any baryon
number produced at temperature $T$ above the electroweak phase
transition temperature will, in general, be destroyed due to {\sl
sphaleron} processes which are in thermal equilibrium for $T>T_{\rm
EW}$. Clearly, the baryon asymmetry can be regenerated at a
temperature around $T_{\rm EW}$, when sphaleron processes fall out of
equilibrium, and this is the so-called {\sl electroweak
baryogenesis}~\cite{ewb1,ewb2}. In the standard electroweak model the
phase transition is too wealkly first-order and the $CP-$violation is
too small to explain the observed baryon asymmetry. The situation
slightly improves in the minimal supersymmetric extension of the
standard model. An alternative to a first-order phase transition can
be obtained by considering the evolution of a defects network, where
the moving topological defects act as expanding bubble walls. There is
a number of such scenarios where cosmic strings have been taken as the
defects~\cite{bs1,bs2,bs3,bs4}.

\subsection{Cosmic strings and ultrahigh energy cosmic rays}
Topological defects in general, and cosmic strings in particular, can
produce high energy particles and contribute to the spectrum of cosmic
rays.  However, what is indeed very interesting is the
possibility~\cite{sigl94} that vacuum defects can explain ultrahigh
energy cosmic rays~\cite{bs99} with energies higher or approximately
equal to $10^{11} {\rm GeV}$, which are in fact hard to explain by
some other more standard astrophysical mechanisms.  Topological
defects produce superheavy Higgs and gauge particles which in their
turn decay into light particles. Defects can easily give particles
with ultrahigh energies and what remains is to to explain the observed
fluxes.

\par Another challenge is to explain the absence of the {\sl
Greisen-Zatsepin-Kuz'min} (GZK) {\sl cutoff}~\cite{g66,zk66} in the
observed cosmic ray spectrum. More precisely, this is the following
puzzle: most models based on reasonable astrophysical assumptions
indicate a likely maximum value for the energy of any kind of emitted
particle of at most approximately $6\times 10^{10}{\rm
GeV}$~\cite{sigl94}.  However, pion production on the background
microwave photons makes it impossible for a particle to propagate with
such energy on scales much larger than a few tens of Mpc, which is
known as the GZK cutoff, and it is expected in all models with a
uniform distribution of sources. On the orther hand, from the
observational point of view, there is evidence for a cosmic ray at an
energy of $(3.2\pm 0.6)10^{11}{\rm GeV}$ and there is also evidence
for showers above $10^{11}{\rm GeV}$, and well above the GZK
cutoff~\cite{linsley63,lawrenceetal91}.

\par Within the context of topological defects, various mechanisms
have been suggested in order to explain the production of ultrahigh
energy particles. For example two mechanisms that have been suggested
are the annihilation of monopole-antimonopole bound
states~\cite{mam1,mam2,mam3} and the annihilation of overlapping
string segments near a cusp~\cite{ass1}. However both those mechanisms
face difficulties to meet the observational data. Another class of
models involves hybrid monopole-string defects~\cite{hms1,hms2,hms3}
and the various proposed scenario are rather plausible.  Moreover
another successful proposition states that decay of metastable vortons
can act as a source of cosmic rays~\cite{mv1}. Finally, there is
another interesting scenario which suggests that utrahigh energy
particles could themselves be topological defects. For example,
magnetic monopoles can be accelerated to energies of the order of
$10^{11}{\rm GeV}$ in galactic magnetic fields~\cite{mm1,mm2,mm3}.  In
this approach, another possibility is that ultrahigh energy cosmic
rays might be the bound states of very massive particles in
vortons~\cite{bp}.  
If we demand that vortons are the candidates for the few $10^{11}{\rm
GeV}$ events, through interaction with atmospheric protons, then the
energy scale $m$ at which strings are formed was found~\cite{bp} to be
$10^9{\rm GeV}$, which is exactly the upper limit on $m$ so that if
the vortons are stable, we can avoid cosmological mass excess (a
rather remarquable coincidence!).  
Acceleration is performed by kicking the vortons with the high
electrostatic fields,
while propagation in the
intergalactic medium is done almost without any collision.
There is no reason for a GZK cutoff and vortons can come from a
small redshift.

\subsection{Cosmic strings as the origin of gamma ray bursts}
As cosmic strings move through cosmic
magnetic fields, they develop electric currents.  It is therefore
natural to expect that oscillating loops of superconducting strings,
emit short bursts of highly beamed electromagnetic radiation, as well
as high-energy particles~\cite{vv87,spg87}.  It was thus
proposed~\cite{bps87,p88} that gamma ray bursts (GRB) could be
produced at cusps of superconducting strings.

\par The original model~\cite{bps87,p88} had assumed that the bursts
originate at very high redshifts 
with GRB photons produced either directly or in electromagnetic
cascades developing due to interaction with the microwave background.
The existence of a strong primordial magnetic field was
required in order to generate the string currents.  However, this
model failed to meet the data.

\par Following basically the same idea, another approach has been more
recently investigated~\cite{bhv00}, where the main mechanism leading to
GRB radiation is the following: low-frequency electromagnetic
radiation from a cusp loses its energy by accelerating particles of
the plasma to very large Lorentz factors.  The particles are beamed
and give rise to a hydrodynamical flow in the surrounding gas,
terminated by a shock.  This model assumes that cosmic magnetic fields
were generated at moderate redshifts 
and then they
remain frozen in the extragalactic plasma.
A single free parameter, the string symmetry breaking scale $\eta\sim
10^{14} {\rm GeV}$, explains the GRB rate, the duration and fluence,
defined as the total energy per unit area of the detector, as well as
the observed ranges of these quantities. 
This model also predicts that GRBs are accompanied by strong bursts of
gravitational radiation ({\sl see} next section).  
These gravitational waves bursts are much stronger than those expected
from more conventional sources and they should be detectable by the
planned LIGO, VIRGO, and LISA detectors.

\section{Gauge cosmic strings and gravity}
Gravitational interactions of strings are characterized by the
dimensionless parameter 
\be 
G\mu\sim (\eta/M_{\rm Pl})^2~, 
\ee 
where $G$ is Newton's constant, $\mu$ is the string mass per unit
length, $\eta$ is the string symmetry breaking scale, and $M_{\rm Pl}$
denotes the Planck mass.

\par Around a straight string the metric is that of a conical space,
which is an almost flat space with a wedge $\Delta=8\pi G\mu$ removed
and the two faces of the wedge being identified. As a consequence of
the conical geometry, light sources behind the string will form double
images~\cite{di1,di2} and therefore strings will act as gravitational
lenses.

\par The relativistic motion of oscillating string loops results to a
strong emission of gravitational radiation.  Oscillating loops of
cosmic strings lose their energy by emitting gravity waves at the
rate~\cite{vv85} \be \dot E=\Gamma G \mu^2~, \ee where the numerical
factor $\Gamma$ only depends  on the shape and the trajectory of the
loop, but it is independent of its length. The typical value for
$\Gamma$ is $50-100$. Thus, the lifetime $\tau$ of a loop of length
$L$ is approximately \be \tau\sim {M\over \dot E}\sim {L\over \Gamma
G\mu}~, \ee where $M=\mu L$ is the mass of the loop.

\par Gravitational radiation from an oscillating loop also carries
away momentum. The rate of momentum radiation is~\cite{vv85} \be |\dot
{\bf P}|=\Gamma_{\rm P}G\mu^2~,\ee with $\Gamma_{\rm P}\sim
10$. finally the rate of angular momentum radiation from loops is
found to be ~\cite{rd89} \be |\dot {\bf J}|=\Gamma_{\rm J}G\mu^2
L~,\ee with $\Gamma_{\rm J}\sim 10$.

\par Gravitational waves emitted by oscillating string loops at
different epochs produce a stochastic gravitational wave background.
The power spectrum of this background covers a large range of
frequencies and over much of this range there is equal logarithmic
frequnecy interval~\cite{vil81}.

\par Long strings are not straight but they have a substantial
small-scale structure in the form of kinks and wiggles on scales below
the characteristic scale of the network~\cite{sa,sv}.  Thus, long
wiggly strings have also gravitational effects which are qualitatively
different from those produced by a straight string. The calculation of
the radiation power from wiggly strings was first presented in
Ref.~\cite{sakel}, where it was applied to the gravitational
radiation from a helicoidal standing wave on a string.

\par Estimating the gravitational wave background from a network of
cosmic strings~\cite{cbs}, one can constrain the string energy scale
$G\mu$ using pulsar timing measurements, as well as
nucleosynthesis. It was found that the result depends on the loop
spectrum, while it is insensitive to the loop size. If the radiation
of the string loops is dominated by the kinks, then the constraint on
the parameter $G\mu$ was found to be: \be G\mu\leq 5.4 (\pm 1.1)\times
10^{-6}~.  \ee Future generations of gravity wave detectors will be
able to probe the predicted spectrum of cosmic strings with $G\mu\sim
10^{-6}$.

\section{Topological defects and brane world cosmology}
Spacetime may have more than four dimensions, with extra coordinates
being unobservable at available energies. 
A first possibility arises in Kaluza-Klein type theories, where the
$D$-dim metric has te form:
\begin{equation}
ds^2=g_{\mu\nu}(x^\mu){\rm d}x^\mu{\rm d}x^\nu
-\gamma_{ab}(x^a){\rm d}x^a{\rm d}x^b~,
\end{equation}
where $g_{\mu\nu}$ is the metric of our four-dimensional world,
$\gamma_{ab}$ is the metric associated with $D-4$ small compact extra
dimensions.  An alternative to Kaluza-Klein compactification proposes
the four dimensions of our world to be identified with the internal
space of topological defects embedded in a higher-dimensional
spacetime (eg., a domain wall in 5d, string in 6d, monopole in 7d,
instanton in 8d, etc), with non-compact extra
dimensions~\cite{rs83,v99,a00}.

\par Then one faces the natural question of how gauge fields and
gravity can be localized on topological defects, in order to make the
whole construction realistic.  The matter fields are localized on the
brane because of the specific dynamics of solitons in string theory
($D$-branes)~\cite{p55}.  Moreover, the gravity of a domain wall in
five-dimensional anti-de Sitter  has a four-dimensional character
for the particles living on the brane, provided the domain wall
tension is fine tuned to a bulk cosmological constant~\cite{rs99}.  In
six dimensions there is a metric solution where gravity is localized
on a four-dimensional singular string-like defect. No tunning of the
bulk cosmological constant to the brane tension is required in order
to cancel the four-dimensional cosmological constant~\cite{gs}.  In
the case of more than two extra dimensions $(n)$, we will distinguish
between local and global defects. It was found~\cite{grs}, that for
strictly (meaning that the stress-energy tensor of the defect is zero
outside the core) local defects, in contrast to the case of one or two
extra dim, there are no solutions that localize gravity when $n\geq
3$.  On the other hand, global defects can lead to the localization of
gravity~\cite{grs}. In the particular case of monopole type
configurations, the introduction of a bulk {\sl hedgehog} magnetic
field leads to a regular geometry and localizes gravity on the
three-brane with either positive, zero or negative bulk cosmological
constant~\cite{grs}.

\section{Conclusions}
Topological defects in general, and cosmic strings in particular,
present a very active subject of modern cosmology. These objects arise
in a wide class of elementary paricle physics models and they are
inevitably formed during phase transitions in the early universe.
They  present a fruitful interplay between cosmology and high
energy physics. Lately defects gained even more interest since one can
{\sl study} various defects aspects in condensed matter experiments.

\par In these lectures I briefly presented topological defects
from the point of view of field theories and I then discussed some of
their cosmological/astrophysical consequences. 

\par Topological defects are ruled out as the main mechanism for the
observed large-scale structure and for the anisotropies of the cosmic
microwave backround radiation. Nevertheless, cosmic strings can still
play a non-negligible r\^ ole in this aspect.  The task we are facing
now is to use this knowledge (combining the predictions of the models
with defects together with the observational data) to choose an
acceptable inflationary scenario, and also to constrain particle
physics models. In addition, topological defects may have played a r\^
ole in some other physical/astrophysical questions which I have
briefly discussed here.

\section*{Acknowledgments}
It is a pleasure to thank the organizers of the NATO ASI / COSLAB
(ESF) School ``Patterns of symmetry breaking'', for the
invitation to deliver these lectures, as well as for succeding in
creating a friendly and nice atmosphere during the whole school in
Cracow. I would also like to thank all my colleagues, with whom I
collaborated on some of the issues I presented here.

{\small


\begin{thebibliography}{99}
\bibitem{review-td} Vilenkin, A., and Shellard, P.E.S. (2000) Cosmic
Strings and Other Topological Defects, {\it Cambridge University
Press}.
\bibitem{hk} Hindmarsh, M.B., and Kibble, T.W.B. (1995) {\it Rept.\
Prog.\ Phys.}  {\bf 58}, p.~477.
\bibitem{nov} Nielsen, H., and Olesen, P. (1973) {\it Nucl.\ Phys.}\
{\bf B61}, p.~45.
\bibitem{ve82} Vilenkin A., and Everett, A.E. (1982) {\it Phys.\ Rev.\
Lett.}\ {\bf 48}, p.~1867.
\bibitem{vor} Shafi Q., and Vilenkin A. (1984) {\it Phys.\ Rev.}\
{\bf D29}, p.~1870.  
\bibitem{hel1} Donnelly, R.J. (1991) Quantized Vortices in Helium II,
{\it Cambridge University Press}.
\bibitem{hel2} Davis, R.L., and Shellard, E.P.S. (1989) {\it Phys.\
Rev.\ Lett.}\ {\bf 63}, p.~2021.
\bibitem{kibble76} Kibble, T.W.B. (1976) {\it J.\ Phys.}\ {\bf A9},
p.~1387.
\bibitem{zurek1} Zurek, W.H. (1985) {\it Nature} {\bf 317}, p.~505.
\bibitem{zurek2} Zurek, W.H. (1996) {\it Phys.\ Rep.}\ {\bf 276},
p.~177.
\bibitem{kibble} Kibble, T.W.B. (1980) {\it Phys.\ Rep.}\ {\bf 67},
p.~183.
\bibitem{1pt1} Borrill, J., Kibble, T.W.B., Vachaspati, T., and
Vilenkin, A. (1995) {\it Phys.\ Rev.}\ {\bf D52}, p.~1934.
\bibitem{1pt2} Ferrera, A. (1999) {\it Phys.\ Rev.}\ {\bf D59},
p.~123503.
\bibitem{1pt3} de Laix, A., and Vachaspati, T. (1999) {\it Phys.\
Rev.}\ {\bf D59}, p.~045017.
\bibitem{1pt4} Copeland, E.J., Saffin, P.M., and Tornkvist, O. (1999)
{\it Phys.\ Rev.}\ {\bf D61}, p.~105005.
\bibitem{1pt5} Davis, A.-C., and Lilley, M. (2000) {\it Phys.\ Rev.}\
{\bf D61}, p.~043502.
\bibitem{sa} Shellard, E.P.S., and Allen, B (1990) in {\it Formation
and Evolution of Cosmic Strings}, Gibbons, G.W., Hawking, S.W., and
Vachaspati, T. eds. (Cambridge University Press).
\bibitem{sv} Sakellariadou, M., and Vilenkin, A. (1990) {\it Phys.\
Rev.}\ {\bf D42}, p.~349.
\bibitem{ack} Austin, D., Copeland, E., and Kibble, T.W.B. (1993) {\it
Phys.\ Rev.}\ {\bf D48}, p.~5594.
\bibitem{vhs} Vincent, G., Hindmarsh, M.B., and Sakellariadou,
M. (1997) {\it Phys.\ Rev.}\ {\bf D56}, p.~637.
\bibitem{bb}
Mather, J.C., et al. (1990) {\it Astrophys.\ J.}\ {\bf 354}, p.~ L37 (1990).
\bibitem{cheng}
Cheng, E.S., et al. (1996) {\it Astrophys.\ J.}\  {\bf 456}, p.~ L71.
\bibitem{cobe1}
Smoot, G.F.,  et al. (1992) {\it Astrophys.\ J.}\ {\bf 396}, p.~ L1.
\bibitem{cobe2}
Bennett, C.L.,  et al. (1996) {\it Astrophys.\ J.}\ {\bf 464}, p.~ L1.
\bibitem{mspain} Sakellariadou, M. (2000) 
in {\it Recent Developments in Gravitation}, {\sl Proceedings of the
Spanish Relativity Meeting, ERE-99} Bilbao (Spain) 7-10 September
1999, Ibanez, J., ed., p.~113.
\bibitem{mrs} Martin, J., Riazuelo, A., and M.\ Sakellariadou (2000)
{\it Phys.\ Rev.}\ {\bf D61}, p.~083518.
\bibitem{gms}
Gangui, A., Martin, J., and Sakellariadou, M. (2002)
{\it Phys.\ Rev.}\ {\bf D66}, p.~083502.
\bibitem{sasaki} Kodama, H., and Sasaki, M. (1984) {\it Prog.\ Theor.\
Phys.\ Suppl.}\ {\bf 78}, p.~1.
\bibitem{RuthReview}
Durrer, R. (1994) {\it Fund.\ of Cosmic Physics} {\bf 15}, p.~209.
\bibitem{rd} Durrer, R., Kunz, M., and Melchiorri, A. (2002) {\it
Phys.\ Rept.}\ {\bf 364}, p.~1.
\bibitem{ruthzhou} 
Durrer, R., and Zhou, Z.H. (1996)  {\it Phys.\ Rev.}\ {\bf D53}, p.~ 5384.
\bibitem{vest} Veeraraghavan, S., and Stebbins, A. (1990) {\it Ap.\
J.}\ {\bf 365}, p.~37.
\bibitem{cor} Vincent, G., Hindmarsh, M.B., and Sakellariadou,
M. (1997) {\it Phys.\ Rev.}\ {\bf D55}, p.~573.
\bibitem{pen} Pen, U.-L., Seljak, U., and Turok, N. (1997) {\it Phys.\
Rev.\ Lett.}\ {\bf 79}, p.~1611.
\bibitem{ram}
Durrer, R., Gangui, A, and Sakellariadou, M. (1996) 
 {\it Phys.\  Rev.\  Lett.}\ {\bf 76}, p.~579.
\bibitem{rma} Durrer, R., Kunz, M, and Melchiorri, A. (1999) {\it
Phys.\ Rev.}\ {\bf D59}, p.~123005.
\bibitem{neil98} Turok, N. Pen, U.-L., and Seljak, U. (1998) {\it
Phys.\ Rev.}\ {\bf D58}, p.~023506.
\bibitem{cs1} Magueijo, J., Albrecht, A., Coulson, D., and Ferreira,
P.  (1996) {\it Phys.\ Rev.\ Lett.}\ {\bf 76}, p.~2617.
\bibitem{cs2} Battye, R.A., Albrecht, A., and Robinson, J. (1998) {\it
Phys.\ Rev.\ Lett.}\ {\bf 80}, p.~4847.
\bibitem{cs3} Avelino, P.P., Shellard, E.P.S., Wu, J.H., and Allen,
B. (1998) {\it Phys.\ Rev.\ Lett.}\ {\bf 81}, p.~2008.
\bibitem{cs4} Avelino, P.P., Caldwell, R.R., and Martins,
C.J.A.R. (1999) {\it Phys.\ Rev.}\ {\bf D59}, p.~123509.
\bibitem{cs5} Contaldi, C., Hindmarsh, M., and Magueijo, J. (1999)
{\it Phys.\ Rev.\ Lett.}\ {\bf 82}, p.~679.
\bibitem{ruthmairi} Durrer, R. and Sakellariadou, M. (1997) {\it
Phys.\ Rev.}\  {\bf D56}, p.~4480.
\bibitem{lw2002} Lyth, and Wands (2002) {\sl Phys.\ Lett.}\ {\bf
B524}, p.~5.
\bibitem{boom} 
C.~B.~Netterfield, et al. (2002) {\it Astrophys.\ J.}\ {\bf 571}, p.~604.
\bibitem{boom1}
P.~de~Bernardis, et al. (2002) {\it Astrophys.\ J.}\ {\bf 564}, p.~559.
\bibitem{max} 
Lee, A.T., et al. (2001) {\it Astrophys.\ J.}\ {\bf 561}, p.~L1.
\bibitem{max1}
Stompor, R., et al. (2001) {\it Astrophys.\ J.}\ {\bf 561}, p.~L7.
\bibitem{dasi}
Halverson, N.W., et al. (2002) {\it Astrophys.\ J.}\ {\bf 568}, p.~38.
\bibitem{dasi1} 
Pryke, C., et al. (2002) {\it Astrophys.\ J.}\ {\bf 558}, p.~46.
\bibitem{jetal} Brandenberger, R., and Martin, J. (2001) {\it Mod.\
Phys.\ Lett.}\ {\bf A16}, p.~999.
\bibitem{ll} 
Liddle, A.R., Lyth, D.H. (1993) {\it Phys.\ Rep.}\ {\bf 231}, p.~1.
\bibitem{linde2}
Kofman, L.A., and Linde, A.D. (1997) {\it Nucl.\ Phys.\ B} {\bf 282}, p.~555.
\bibitem{linde3}
Linde, A.D., and Riotto, A. (1997) {\it Phys.\ Rev.}\ {\bf D56}, p.~1841.
\bibitem{lr}
Lyth, D.H., and Riotto, A. (1999) {\it Phys.\ Rep.}\ {\bf 314}, p.~1.
\bibitem{bprs}
Bouchet, F.R., Peter, P.,  Riazuelo, A., and Sakellariadou, M. (2001)
{\it Phys.\ Rev.}\ {\bf D65}, p.~021301(R). 
\bibitem{rj96} Jeannerot, R. (1996) {\it Phys.\ Rev.}\ {\bf D53}, p.~5426.
\bibitem{rj97} Jeannerot, R. (1997) {\it Phys.\ Rev.}\ {\bf D56}, p.~6205.
\bibitem{witten85a} Witten, E. (1985) {\it Nucl.\ Phys.}\ B{\bf 249},
p.~557.
\bibitem{vort1} Davis, R.L., and Shellard, E.P.S. (1988) {\it Phys.\
Lett.}\ {\bf B329}, p.~485.
\bibitem{vort2} Davis, R.L., and Shellard, E.P.S. (1988) {\it Nucl.\ Phys.} 
{\bf B323}, p.~209.
\bibitem{betal02} Brandenberger, R., Carter, B., and Davis,
A.-C. (2002) {\it Phys.\ Lett.}\ {\bf B534}, p.~1.
\bibitem{Sakharov 1967} Sakharov, A.D. (1967) {\it JETP Lett.}\ {\bf
5}, p.~24.
\bibitem{bdh91} Prokopec, T., Brandenberger,R., Davis, A.C., and
Trodden, M. (1996) {\it Phys.\ Lett.}\ {\bf B384}, p.~175.
\bibitem{ewb1} Rubakov, V.A., and Shaposhnikov, M.E. (1996) {\it Phys.\ Usp.}\
{\bf 39}, p.~461.
\bibitem{ewb2} Riotto, A., and Trodden, M. (1999) 
{\it Ann.\ Rev.\ Nucl.\ Part.\ Sci.}\ {\bf 49}
\bibitem{bs1} Brandenberger, R., Davis, A.-C., and Trodden, M. (1994)
{\it Phys.\ Lett.}\ B{\bf 335}, p.~123.
\bibitem{bs2} Trodden, M., Davis, A.-C., and Brandenberger, R. (1995)
{\it Phys.\ Lett.}\ B{\bf 349}, p.~131.
\bibitem{bs3} Brandenberger, R., Davis, A.-C., Prokopek, T.,and
  Trodden, M.  (1996) {\it Phys.\ Rev.}\ D{\bf 53}, p.~4257.
\bibitem{bs4} Cline, J., Espinosa, J., Moore, G.D., and Riotto, A. (1999)
{\it Phys.\ Rev.}\ D{\bf 59}, p.~065014.
\bibitem{sigl94} Sigl, G., Schramm, D., and Bhattacharjee, P. (1994)
{\it Astropart.\ Phys.}\ {\bf 2}, p.~401.
\bibitem{bs99}  Bhattacharjee, P., and Sigl (2000)
 {\it Phys.\ Rept.}\ {\bf 327}, p.~109.
\bibitem{g66} Greisen, K. (1966) {\it Phys.\ Rev.\ Lett.}\ {\bf 16}, p.~748.
\bibitem{zk66} Zatsepin, G.T., and Kuz'min V.A. (1966) {\it JETP
Lett.} {\bf 4}, p.~78.
\bibitem{linsley63} Linsley, K (1963) {\it Phys.\ Rev.\ Lett.}\ {\bf
10}, p.~146.
\bibitem{lawrenceetal91} Lawrence, M.A., Reid, R.J.O., and Watson,
A.A. (1991) {\it J.\ Phys.}\ {\bf G17}, p.~733.
\bibitem{mam1} Hill, C.T. (1983) {\it Nucl.\ Phys.}\ {\bf B224}, p.~469.
\bibitem{mam2} Bhattacharjee, P. and Sigl, G. (1995) {\it Phys.\
Rev.}\ {\bf D51}, p.~4079.
\bibitem{mam3} Blanco-Pillado, J.J., and Olum, K.D. (1999) {\it Phys.\
Rev.}\ {\bf D60}, p.~083001
\bibitem{ass1} Olum, K.D., and Blanco-Pillado, J.J. (1999) {\it Phys.\
Rev.}\ D{\bf 60}, p.~023503.
\bibitem{hms1} Berezinsky, V., Martin, X., and Vilenkin, A. (1997)
{\it Phys.\ Rev.}\ D{\bf 56}, p.~2024.
\bibitem{hms2} Berezinsky, V., and Vilenkin, A. (1997)
{\it Phys.\ Rev.\ Lett.}\ {\bf 79}, p.~5202.
\bibitem{hms3} Berezinsky, V., Blasi, P., and Vilenkin, A. (1998)
{\it Phys.\ Rev.}\ D{\bf 58}, p.~103515.
\bibitem{mv1} Masperi, L., and Silva, G. (1998) {\it Astrop.\ Phys.}\
{\bf 8}, p.~173.
\bibitem{mm1} Weiler, T.J., and Kephart, T.W. (1996) {\it Nucl.\
Phys.\ Proc.\ Suppl.}\ {\bf 51B}, p.~218.
\bibitem{mm2} Huguet, E., and Peter, P. (2000) {\it Astropart.\ Phys.}\
{\bf 12}, p.~277.
\bibitem{mm3} Wick, S.D., Kephart, T.W., Weiler, T.J., and Biermann,
P.L. (2000) {\it Astropart.\ Phys.}\ ({\tt astro-ph/0001233}).
\bibitem{bp} Bonazzola, S., and Peter, P. (1997) {\it Astropart.\
Phys.}\ {\bf 7}, p.~161.
\bibitem{vv87} Vilenkin, A., and Vacgaspati, T, (1987) {\it Phys.\
Rev.\ Lett.}\ {\bf 58}, p.~1041.
\bibitem{spg87} Spergel, D.N., Piran, T., and Goodman, J. (1987) {\it
Nucl.\ Phys.}\ B{\bf 291}, p.~847.
\bibitem{bps87} Babul, A., Paczynski, B., and Spergel, D.N. (1987) {\it
Ap.\ J.\ Lett.}\ {\bf 316}, p.~L49.
\bibitem{p88} Paczynski, B. (1988) {\it Ap.\ J.}\ {\bf 335}, p.~525.
\bibitem{bhv00} Berezinsky, V., Hnatyk, B., and Vilenkin, A. (2000)
({\tt astro-ph/0001213}).
\bibitem{di1} Vilenkin, A. (1981) {\it Phys.\ Rev.}\ {\bf D23}, p.~852.
\bibitem{di2} Gott, J.R. (1985) {\it Ap.\ J.}\ {\bf 288}, p.~422.
\bibitem{vv85} Vachaspati, T., and Vilenkin, A. (1985) {\it Phys.\
Rev.}\ {\bf D31}, p.~3052.
\bibitem{rd89} Durrer, R. (1989) {\it Nucl.\ Phys.}\ {\bf B328},
p.~238.
\bibitem{vil81} Vilenkin, A. (1981) {\it Phys.\ Lett.}\ {\bf 107B},
p.~47.
\bibitem{sakel} Sakellariadou, M. (1990)  {\it Phys.\
Rev.}\ {\bf D42}, p.~354.
\bibitem{cbs} Caldwell, R.R., Battye, R.A., and Shellard,
E.P.S. (1996) {\it Phys.\ Rev.}\ {\bf D54}, p.~7146.
\bibitem{rs83} Rubakov, V.A., and Shaposhnikov, M.E. (1983) {\it
Phys.\ Lett.}\ {\bf B 125}, p.~139.
\bibitem{v99} Visser, M. (1985) {\it Phys.\ Lett.}\ {\bf B 159}, p.~22.
\bibitem{a00} Akama, K. (2000) in {\it Proceedings of the Symposium
on Gauge Theory and Gravitation},  Kikkawa, K., Nakanishi, N., and 
Nariai, H., eds.
\bibitem{p55} Polchinski, J. (1995)  {\it Phys.\ Rev.\
Lett.}\ {\bf 75}, p.~4724.
\bibitem{rs99} Randall, L., and Sundrum, R. (1999) {\it Phys.\ Rev.\
Lett.}\ {\bf 83}, p.~4690.
\bibitem{gs} Gherghetta, T., and Shaposhnikov, M. (2000) {\it Phys.\
Rev.\ Lett.}\ {\bf 85}, p.~240.
\bibitem{grs} Gherghetta, T., Roessl, E., and Shaposhnikov, M. (2000)
{\it Phys.\ Lett.}\ {\bf B491}, p.~353.

\end{thebibliography}
\end{document}